\documentclass[aps,pra,showpacs]{revtex4}
\usepackage{amsfonts}
\usepackage{amssymb}
\usepackage{bm}

\begin{document}


\title{Phase states for a three-level atom
interacting with quantum fields}

\author{A. B. Klimov}
\affiliation{Departamento de F\'{\i}sica,
Universidad de Guadalajara,
Revoluci\'on 1500,
44420~Guadalajara, Jalisco,
Mexico}

\author{L. L. S\'anchez-Soto, J. Delgado,
and E. C. Yustas}
\affiliation{Departamento de \'{O}ptica,
Facultad de Ciencias F\'{\i}sicas,
Universidad Complutense,
28040 Madrid, Spain}

\date{\today}

\begin{abstract}
We introduce phase operators associated with
the algebra su(3), which is the appropriate tool
to describe three-level systems. The rather
unusual properties of this phase are caused by
the small dimension of the system and are
explored in detail. When a three-level atom
interacts with a quantum field in a cavity, a
polynomial deformation of this algebra emerges
in a natural way. We also introduce a polar
decomposition of the atom-field relative
amplitudes that leads to a Hermitian
relative-phase operator, whose eigenstates
correctly describe the corresponding phase
properties. We claim that this is the natural
variable to deal with quantum interference
effects in atom-field interactions. We find
the probability distribution for this variable
and study its time evolution in some special cases.
\end{abstract}


\maketitle

\section{Introduction}

Atomic coherence is essential for a proper understanding
of many important effects appearing in the response of an
atomic system to strong laser radiation. Perhaps the
Mollow spectrum~\cite{Mollow72} for a strongly driven
two-level atom has been the first milestone in
quantum optics where atomic coherence plays a major
role.

Once the population dynamics in two-level systems was
understood, the interest in multilevel (and, especially,
three-level) atoms became soon apparent. A significant
amount of research has thus been devoted to their study
and many effects based on quantum coherence such as
spectroscopic dark states~\cite{Orriols76}, electron
shelving~\cite{Hegerfeldt92}, narrow spectral
lines~\cite{narrow}, pulse matching~\cite{Cook79},
and anti-intuitive excitation~\cite{anti}, have been
highlighted. These nonclassical features have an
enormous variety of interesting and nontrivial
consequences, including electromagnetically induced
transparency~\cite{EIT}, lasing without
inversion~\cite{LWI}, state-selective molecular
excitation~\cite{STIRAP}, and demonstrations of slow
light~\cite{slowlight} and fast light~\cite{fastlight}
to mention only a few examples.

For a three-level atom coupled to one or two
quantized cavity field modes both the atomic
dynamics and the statistical properties of the
reemitted field have been discussed in
detail~\cite{Eberly85} showing interesting
phenomena, the most outstanding being, perhaps,
the existence of collapses and revivals. The phase
properties of this model have also been studied,
mainly using the Pegg-Barnett formalism for the field,
and the connection between revivals and phase has
been put forward~\cite{Aliskenderov91}.

In spite of these achievements, even a cursory look
at any classical model of atom-field interaction
immediately reveals that the natural way of
understanding the resonant behavior is in
terms of the relative phase between the
field and the atomic dipole. It is also
clear that this relative phase is basic
in the treatment of concepts such as
interference or coherence that are
ubiquituous in classical optics.

However, when one tries to extend this variable
into the quantum domain one is immediately
faced with two serious problems: first, there
is no precise prescription to deal with the phase
of a three-level atom~\cite{Sanders01}; second,
we lack a satisfactory description of the
atom-field interaction in terms of relative phase.
The main goal of this paper is precisely to remedy
these drawbacks by proposing a comprehensive theory
of the relative phase for three-level atoms
interacting with quantum fields.

Phase for three-level systems has been handled by
invoking fuzzy concepts such as the phase of the
associated wavefunction~\cite{Buckle86}. We
emphasize that these notions, though well
established in the classical limit, are not easily
extrapolated into the realm of the quantum world.
Since  phase is considered to be a physical
property, in the orthodox picture of quantum
mechanics it must be associated with a selfadjoint
operator or at least with a family of positive
operator-valued measures~\cite{Helstrom76}. In this
spirit, phase operators for the algebra su(2),
which describes two-level systems, have been
previously worked out~\cite{Levy73,Vourdas90,Ellinas90}.
Since su(3) is the natural arena in which to deal with
three-level atoms, we propose here an extension
enabling us to introduce phase operators for these
systems with a clear physical picture.

Concerning the aforementioned second problem, we
stress that an operator describing the relative
phase between field and atoms has resisted a quantum
description~\cite{Luis97,Delgado01}. For this reason,
its role has been played by the field
quadratures~\cite{Allen87}, which in many
respects properly account for phase relations
and are free from the difficulties that phase
encounters in the quantum domain.
When focusing on the relative phase between
two subsystems, we think the best way to proceed,
much in the spirit of our previous work on the
subject~\cite{Delgado00}, is to try a polar
decomposition of the quantum amplitudes that
parallels as much as possible the corresponding
classical factorization.

For the relative phase between two quantum field
modes this is quite a straightforward procedure and
leads to a unitary solution~\cite{relpha}. For the
case at hand, this polar decomposition seems to be
more involved, mainly because, unlike for the case of
two harmonic oscillators, the Hamiltonian cannot be
cast in terms of su(3) generators, but rather in terms
of a polynomial deformation of su(3). These nonlinear
algebras have been examined very recently in quite
different physical contexts~\cite{deformed}.
Without embarking on mathematical subtleties,
we shall exploit these previous results to perform the
polar decomposition in an elegant way, obtaining
a bona fide Hermitian operator representing
the relative phase we wish to examine.

It is our firm belief that this Hermitian operator and
its eigenstates should be a fundamental tool for
examining all the issues related with quantum
interference between atomic pathways. This leads
to coherent effects as those mentioned earlier.
In this paper we use this operator to introduce
its associated probability distribution and to
explain some relevant dynamical features of the
model. Nevertheless we stress that given the
variety of phenomena of both conceptual and
practical importance arising in this field,
it is outside the scope of this paper to
provide a full account of them. Rather, our
goal is to emphasize the role that the relative
phase plays in atom-field interactions, a
role that has previously gone almost unnoticed.

\section{Phase for three-level systems}

We wish to explore in some detail the phase
properties of a three-level system. To be specific
we shall consider a $\Lambda$ configuration, as
shown in Fig.~1, with energy levels $\omega_1 <
\omega_2 < \omega_3$ (we shall use throughout all
this paper units $\hbar = 1$) and with allowed
dipole transitions $1 \leftrightarrow 3$ and
$2 \leftrightarrow 3$, but not $1 \leftrightarrow 2$.

In order to describe the kinematics of this system
one must take into account that the natural generalization
of the Bloch vector~\cite{Allen87} (which is essential
in understanding the behavior of a two-level system)
comprises now the nine operators
\begin{equation}
\label{S}
\hat{S}^{ij} = | j \rangle \langle i | ,
\end{equation}
where $| i \rangle$ denotes the eigenstate of the
$i$th atomic level.  One can easily
check that they satisfy
\begin{equation}
\label{ccr3}
[\hat{S}^{ij}, \hat{S}^{kl} ] =  \delta_{il} \hat{S}^{kj}
-\delta_{kj} \hat{S}^{il}  ,
\end{equation}
which correspond to the commutation relations of
the algebra u(3)~\cite{Gilmore74}.

Obviously, the three ``diagonal'' operators $\hat{S}^{ii}$
measure level populations, while the ``off-diagonal''
ladder operators $\hat{S}^{ij}$ generate transitions from
level $i$ to level $j$. To emphasize this idea we
define raising and lowering operators by
\begin{equation}
\left \{
\begin{array}{lll}
\hat{S}_+^{ij} & =  \hat{S}^{ij}
\qquad &
\mbox{\textrm{when \ }} j> i, \\
&  & \\
\hat{S}_-^{ji} & = \hat{S}^{ij}
\qquad
&
\mbox{\textrm{when \ }} j < i .
\end{array}
\right .
\end{equation}

Because one has the trivial constraint $ \hat{S}^{11}
+ \hat{S}^{22} + \hat{S}^{33}= \hat{I}$, only two
populations can vary independently. For this reason,
it is customary to introduce two independent traceless
operators
\begin{equation}
\hat{S}_z^{13} = \frac{1}{2}
(\hat{S}^{33} - \hat{S}^{11}) ,
\qquad
\hat{S}_z^{23} = \frac{1}{2}
(\hat{S}^{33} - \hat{S}^{22}) ,
\end{equation}
that measure atomic inversions between the corresponding
levels. The operators $(\hat{S}_\pm^{ij}, \hat{S}_z^{ij})$
($i \ne j)$ turn out to be the eight generators of the
algebra su(3). Note that $(\hat{S}_\pm^{13}, \hat{S}_z^{13})$
and $(\hat{S}_\pm^{23}, \hat{S}_z^{23})$  form two
su(2) subalgebras that physically correspond to
the two dipoles that appear in the two allowed
transitions $1 \leftrightarrow 3$ and
$2 \leftrightarrow 3$. However, these two dipoles
are not independent, since Eq.~(\ref{ccr3}) imposes
\begin{equation}
\label{nlc}
[\hat{S}_+^{13}, \hat{S}_-^{23} ]= - \hat{S}_+^{12} ,
\qquad
[\hat{S}_+^{13}, \hat{S}_+^{23} ] = 0 ,
\end{equation}
which constitutes a highly nontrivial coupling between
them.

It is usual to take for granted the existence of the
phase of the atomic dipoles, mainly because classically
one has a clear picture of their meaning. However,
strictly speaking, we do not have any prescription to
deal with this variable at the quantum level. Nevertheless,
as anticipated in the Introduction, one could expect to
get a formalism very close to the one developed recently
for two-level systems.

For definiteness let us first focus on the transition
$1 \leftrightarrow 3$. If only these two levels were
involved, such a transition would be described by a
superposition of the form
\begin{equation}
\label{qubit}
| \Psi \rangle = \sin(\vartheta/2) \ | 1 \rangle +
e^{i \varphi} \ \cos(\vartheta/2) \ | 3 \rangle .
\end{equation}
This corresponds to a 1/2 angular-momentum system.
For this case~\cite{Cohen90} the Bloch vector is
$\hat{\mathbf{S}} = \hat{\bm{\sigma}}/2$,
$\hat{\bm{\sigma}}$ being the Pauli matrices,
and it is easy to work out that the mean values
$s_j = \langle \Psi | \hat{S}_j | \Psi \rangle$ are
given by
\begin{eqnarray}
s_x & = & \sin \vartheta \ \cos \varphi , \nonumber \\
s_y & = & \sin \vartheta \ \sin \varphi , \nonumber \\
s_z & = & \cos \vartheta .
\end{eqnarray}
This would support the naive belief that, when viewed in
the Bloch sphere $s_x^2 + s_y^2 + s_z^2 =1$, the
parameter $\varphi$ is the phase angle associated
with the atomic dipole $1 \leftrightarrow 3$
and is canonically conjugate to $s_z$~\cite{Luis00}.
In consequence, there is a widespread usage of dealing
with this dipole phase $\varphi$ as a state parameter
instead of a quantum variable. To gain further insight
into this point, let us note that
\begin{equation}
\label{s-}
s_- =  s_x - i s_y =
\sin \vartheta \ e^{i \varphi} ,
\end{equation}
so it is clear that the parameter ``dipole phase" can
be obtained through the decomposition of $s_-$ in
terms of modulus and phase. Obviously, it is tempting
to pursue this simple picture by taking into account that,
at the operator level, the equivalent to the decomposition
in terms of modulus and phase is a polar decomposition.
Thus, it seems appropriate to define the operator
counterpart of Eq.~(\ref{s-}) in the three-level case
we are considering as
\begin{equation}
\label{polar}
\hat{S}^{13}_- =  \sqrt{\hat{S}_-^{13} \hat{S}_+^{13}} \
\hat{E}^{13}_\varphi .
\end{equation}
Here $\hat{E}^{13}_\varphi = \exp(i \hat{\varphi}^{13})$,
where $\hat{\varphi}^{13}$ is the Hermitian operator
representing the phase of the transition.

One can work out that the unitary solution of
Eq.~(\ref{polar}) may be written as
\begin{equation}
\hat{E}^{13}_\varphi = | 1 \rangle \langle 3 |
+ e^{i \varphi_0^\prime} \ | 3 \rangle \langle 1 |
- e^{-i \varphi_0^\prime} \ | 2 \rangle \langle 2 | ,
\end{equation}
where the undefined factor $e^{ i \varphi_0^\prime}$
appears due to the unitarity requirement of
$\hat{E}^{13}_\varphi$. The main features of
this operator are largely independent of
$\varphi_0^\prime$, but for the sake of concreteness,
we can make a definite choice. For
example~\cite{Luis97}, if we assume a state
such as the linear superposition (\ref{qubit}),
the complex conjugation of the wavefunction
should reverse the sign of $\hat{\varphi}^{13}$,
which immediately leads to the condition
$e^{i \varphi_0^\prime}= -1$. We conclude
then that the operator we are looking for
can be represented as
\begin{equation}
\hat{E}^{13}_\varphi = | 1 \rangle \langle 3 |
- | 3 \rangle \langle 1 |
+  | 2 \rangle \langle 2 | .
\end{equation}
The associated eigenstates are
\begin{eqnarray}
\label{eigenv13}
| \varphi^{13}_0 \rangle & = & | 2 \rangle ,
\nonumber \\
& & \\
| \varphi^{13}_\pm \rangle & = &
\frac{1}{\sqrt{2}}
( | 3 \rangle \pm i | 1 \rangle ) ,
\nonumber
\end{eqnarray}
and the  eigenvalues of $\hat{\varphi}^{13}$
are 0 and $\pm \pi/2$, respectively. This is
a remarkable result. It shows that the
eigenvectors $| \varphi^{13}_\pm \rangle$ look like
the standard ones for a quantum dipole or a spin 1/2.
However, the ``spectator'' level $| 2 \rangle$
(i.e., it apparently does not take part in
the transition) is an eigenstate of this operator,
which introduces drastic changes in the
dynamics. In other words, the phase of the transition
$1 \leftrightarrow 3$ ``feels" the state $| 2 \rangle$.

Perhaps, the most striking feature of this theory is
that a measurement of the phase of the transition
gives only three possible values: 0 and $\pm \pi/2$.
While this kind of statement seems rather reasonable
when dealing with spin systems, they are scarcely
recognized when dealing with atoms.

An analogous reasoning for the transition
$2 \leftrightarrow 3$ gives the corresponding
operator $\hat{E}^{23}_\varphi$
\begin{equation}
\hat{E}^{23}_\varphi = | 2 \rangle \langle 3 |
- | 3 \rangle \langle 2 |
+  | 1 \rangle \langle 1 | ,
\end{equation}
with eigenvectors
\begin{eqnarray}
| \varphi^{23}_0 \rangle & = & | 1 \rangle ,
\nonumber \\
& & \\
| \varphi^{23}_\pm \rangle & = &
\frac{1}{\sqrt{2}}
( | 3 \rangle \pm i | 2 \rangle ) ,
\nonumber
\end{eqnarray}
and the same eigenvalues as before.

Note that
\begin{equation}
\hat{E}^{13}_\varphi
\hat{E}^{23}_\varphi{}^\dagger
\neq \hat{E}^{12}_\varphi ,
\end{equation}
which clearly displays the quantum nature
of this phase.

To any smooth function $F(\varphi^{ij})$ of
the dipole phase $\varphi^{ij}$ we can associate the
operator
\begin{equation}
F(\hat{\varphi}^{ij}) = \sum_{r = 0, \pm}
| \varphi^{ij}_r \rangle \
F (\varphi^{ij}_r) \
 \langle \varphi^{ij}_r | ,
\end{equation}
where the sum runs over the eigenvalues 0 and $\pm \pi/2$.
The mean value of this operator function can be computed
as
\begin{equation}
\label{Fmean}
\langle F(\hat{\varphi}^{ij}) \rangle =
\sum_r F (\varphi^{ij}_r) P(\varphi^{ij}_r) ,
\end{equation}
where $ P(\varphi^{ij}_r)$ is the probability distribution
\begin{equation}
\label{Ptheta}
P(\varphi^{ij}_r) = \mathrm{Tr}
\left [ \hat{\varrho} \
| \varphi^{ij}_r \rangle
\langle \varphi^{ij}_r | \right ] ,
\end{equation}
for any state described by the density matrix
$\hat{\varrho}$. We shall make use of these results
in next Section.

\section{Exploring the role of the relative phase}

\subsection{Quantum dynamics of a three-level
atom coupled to a two-mode field}

To maintain the discussion as self contained as possible
we briefly review the basic aspects of the interaction
of a three-level atom with a two-mode field. We shall
merely outline what will be essential for our later
discussion of phase properties. We write the Hamiltonian
for this system as
\begin{equation}
\label{H3l}
\hat{H} = \hat{H}_{\mathrm{A}} +
\hat{H}_{\mathrm{F}} + \hat{V} ,
\end{equation}
where
\begin{eqnarray}
\hat{H}_{\mathrm{A}} & = & \sum_i \omega_i
\hat{S}^{ii} ,
\nonumber \\
\hat{H}_{\mathrm{F}} & = &
\omega_a \hat{a}^\dagger \hat{a} +
\omega_b  \hat{b}^\dagger \hat{b} ,
\nonumber \\
\hat{V} & = & g_a (\hat{a} \hat{S}_+^{13} +
\hat{a}^\dagger \hat{S}_-^{13})
+ g_b (\hat{b} \hat{S}_+^{23} +
\hat{b}^\dagger \hat{S}_-^{23}) .
\end{eqnarray}
Here $\hat{H}_{\mathrm{A}}$ describes the dynamics of
the free atom and  $\hat{H}_{\mathrm{F}}$ represents
the cavity modes of frequency $\omega_a$ and $\omega_b$,
with annihilation operators $\hat{a}$ and $\hat{b}$,
respectively. Finally, in the interaction term $\hat{V}$,
written in the dipole and rotating-wave approximations,
we assume that the allowed transition $1 \leftrightarrow 3$
couples (quasi)resonantly to the mode $a$ and the transition
$2 \leftrightarrow 3$ couples to the mode $b$
with coupling constants $g_a$ and $g_b$ that will
be taken as real numbers.

The bare basis for the total system is
$| i \rangle_{\mathrm{A}} \otimes
| n_a, n_b \rangle_{\mathrm{F}}$,
where $| n_a, n_b \rangle_{\mathrm{F}}$,
is the usual two-mode Fock basis. However,
one can check that the two excitation-number
operators
\begin{equation}
\hat{N}_a  =  \hat{a}^\dagger \hat{a} -
\hat{S}^{11} + 1 ,
\qquad
\hat{N}_b  =  \hat{b}^\dagger \hat{b} -
\hat{S}^{22} + 1 ,
\end{equation}
are conserved quantities. In consequence, we
can rewrite the Hamiltonian (\ref{H3l}) as
\begin{equation}
\hat{H} = \hat{H}_0 + \hat{H}_{\mathrm{int}} ,
\end{equation}
where
\begin{eqnarray}
\hat{H}_0 & = & \omega_a \hat{N}_a +
\omega_b \hat{N}_b +
(\omega_3 - \omega_a - \omega_b ) \hat{I},
\nonumber \\
& & \\
\hat{H}_{\mathrm{int}} & = & -\Delta_a \hat{S}^{11}
-\Delta_b \hat{S}^{22} \nonumber \\
& + & g_a (\hat{a} \hat{S}_+^{13} +
\hat{a}^\dagger \hat{S}_-^{13})
+ g_b (\hat{b} \hat{S}_+^{23} +
\hat{b}^\dagger \hat{S}_-^{23}) .
\nonumber
\end{eqnarray}
The detunings are defined as
\begin{equation}
\Delta_a = \omega_{31} - \omega_a ,
\qquad
\Delta_b = \omega_{32} - \omega_b ,
\end{equation}
with $\omega_{ij} = \omega_i - \omega_j$.
It is a straightforward calculation to check that
\begin{equation}
[ \hat{H}_0 , \hat{H}_{\mathrm{int}}] = 0 .
\end{equation}
Therefore, both the free Hamiltonian $\hat{H}_0$
and the interaction Hamiltonian $\hat{H}_{\mathrm{int}}$
are constants of motion. $\hat{H}_0$ determines the
total energy stored in the system, which is
conserved by the interaction. This allows us
to factor out $\exp(- i \hat{H}_0 t)$ from the
evolution operator and drop it altogether. Thus,
the problem can be reduced to study the restriction
of $\hat{H}_{\mathrm{int}}$ to each subspace
$\mathcal{H}^{(N_a, N_b)}$ with fixed values
of the pair of excitation numbers $(N_a, N_b)$.
In each one of these subspaces
$\mathcal{H}^{(N_a, N_b)}$ there are three basis
vectors that can be written as
\begin{equation}
| i; n_a = N_a-\mu_i, n_b= N_b- \nu_i \rangle  ,
\end{equation}
where the values of $\mu_i$ and $\nu_i$ are defined as
\begin{equation}
(\mu_1, \mu_2, \mu_3) = (0, 1, 1) ,
\qquad
(\nu_1, \nu_2, \nu_3) = (1, 0, 1) .
\end{equation}
Note that when $N_a=1$ and $N_b =0$ or $N_a=0$ and
$N_b =1$ some states may have negative photon
occupation number and must be eliminated. In
the subspace $\mathcal{H}^{(N_a, N_b)}$,
$\hat{H}_{\mathrm{int}}$ is represented  by the
$3 \times 3$ matrix
\begin{equation}
\hat{H}_{\mathrm{int}}^{(N_a, N_b)} =
\left (
\begin{array}{ccc}
0 & g_b \sqrt{N_b} & g_a  \sqrt{N_a} \\
g_b \sqrt{N_b} & - \Delta_b & 0 \\
 g_a  \sqrt{N_a} & 0 & - \Delta_a
\end{array}
\right ) .
\end{equation}

Let us assume that at $t=0$ the atomic
wave function can be written as the coherent
(normalized) superposition
\begin{equation}
| \Psi (0) \rangle_{\mathrm{A}} =
\sum_i c_i \ | i \rangle_{\mathrm{A}} ,
\end{equation}
while the field is in a two-mode coherent state
\begin{equation}
| \Psi (0) \rangle_{\mathrm{F}} =
|\alpha_a, \alpha_b \rangle =
\sum_{n_a, n_b = 0}^\infty
 Q_{n_a} Q_{n_b} \
|n_a, n_b \rangle_{\mathrm{F}} .
\end{equation}
Here $Q_n$ is the Poissonian weighting
factor of the coherent state (with zero phase)
with mean number of photons $\bar{n}$:
\begin{equation}
Q_n = \sqrt{e^{- \bar{n}}
\frac{\bar{n}^n}{n!}} .
\end{equation}

At a later time $t$  the state vector for
the atom-field system in the interaction
picture can be expressed as
\begin{eqnarray}
\label{Psit}
| \Psi (t) \rangle  & = &
\sum_{N_a, N_b = 0}^\infty
\sum_{i, j = 1}^{3}
Q_{N_a - \mu_j} \ Q_{N_b - \nu_j} \ c_j \nonumber \\
& \times & {\mathcal{U}}_{ij}^{(N_a, N_b)} (t) \
|i, N_a - \mu_i , N_b- \nu_i \rangle ,
\end{eqnarray}
where ${\mathcal{U}}_{ij}^{(N_a, N_b)} (t)$
are the matrix elements of the evolution operator
in the subspace $\mathcal{H}^{(N_a, N_b)}$
\begin{equation}
{\mathcal{U}}_{ij}^{(N_a, N_b)} (t)   =
\langle i; N_a-\mu_i, N_b- \nu_i |
\exp [ - i \hat{H}_{\mathrm{int}}^{(N_a, N_b)} t ]
|j; N_a-\mu_j, N_b- \nu_j \rangle ,
\end{equation}
which can be calculated exactly as can be seen, e.g.
in Ref~\cite{Eberly85}. This state describes completely
the system evolution and will be the basis for our
phase analysis in the following.

\subsection{Polar decomposition of the
relative atom-field amplitudes: deformed
su(3) dynamics}

As stated in the Introduction, our goal is to describe
the atom-field relative phase by resorting to a
polar decomposition of the corresponding complex
amplitudes. To this end, let us define the operators
\begin {eqnarray}
& \hat{X}_+^{13} = \hat{a} \ \hat{S}_+^{13},
\qquad
\hat{X}_z^{13} = \hat{S}_z^{13} ; & \nonumber \\
& & \\
& \hat{X}_+^{23} = \hat{b} \ \hat{S}_+^{23},
\qquad
\hat{X}_z^{23} = \hat{S}_z^{23} . &
\nonumber
\end{eqnarray}
These operators satisfy most of the usual su(3)
commutation relations  provided (\ref{nlc})
is recast as
\begin{equation}
[\hat{X}_+^{13}, \hat{X}_-^{23} ]= -
\hat{Y}_+^{12} ,
\end{equation}
where
\begin{equation}
\hat{Y}_+^{12} = - \hat{a} \hat{b}^\dagger
\hat{S}_+^{12} .
\end{equation}
However, some of them must be modified in the following
way
\begin{eqnarray}
\ [ \hat{X}_+^{13}, \hat{X}_-^{13} ]  & = &
\hat{N}_a (1-2 S^{11} - S^{22} ), \nonumber \\
\ [ \hat{X}_+^{23}, \hat{X}_-^{23} ]  & = &
\hat{N}_b (1-2S^{22}-S^{11}), \\
\ [ \hat{Y}_+^{12}, \hat{Y}_-^{12} ]  & = & \hat{N}_a
\hat{N}_b (S^{11}-S^{22}), \nonumber
\end{eqnarray}
which corresponds to a polynomial deformation
of the algebra su(3).  The essential point
for us is that one can develop a theory in
very close analogy with the standard su(3)
algebra.

For simplicity, let us focus first on the
allowed transition $1 \leftrightarrow 3$ and
notice that in every three-dimensional invariant
subspace $\mathcal{H}^{(N_a, N_b)}$ the state
$ |1; N_a, N_b -1 \rangle $ plays the role of a
\textit{vacuum state} since
\begin{equation}
\hat{X}_-^{13} |1; N_a, N_b -1 \rangle = 0 .
\end{equation}
In this subspace the operator $\hat{X}_z^{13}$
is diagonal  and we can work out again a polar
decomposition similar to (\ref{polar}), namely
\begin{equation}
\label{poldef}
\hat{X}_-^{13}  =
\sqrt{\hat{X}_-^{13} \hat{X}_+^{13}} \
\hat{E}^{13}_\Phi ,
\end{equation}
where the operator $\sqrt{\hat{X}_-^{13} \hat{X}_+^{13}}$
is diagonal and
\begin{eqnarray}
& \hat{E}^{13}_\Phi \hat{E}^{13}_\Phi{}^\dagger =
\hat{E}^{13}_\Phi{}^\dagger
\hat{E}^{13}_\Phi = \hat{I} , & \nonumber \\
& & \\
& [\hat{E}^{13}_\Phi, \hat{N}_a ] =
[ \hat{E}^{13}_\Phi, \hat{N}_b] = 0 . &
\nonumber
\end{eqnarray}
The first equation ensures that the operator
$\hat{E}^{13}_\Phi= \exp(i \hat{\Phi}^{13})$,
representing the exponential of the relative
phase between the field and the dipole
$1 \leftrightarrow 3$, is unitary. The second one
guarantees that we may study its restriction to each
invariant subspace $\mathcal{H}^{(N_a, N_b)}$.

Much in the same way as we did in Sec. II,
the operator $\hat{E}^{13}_\Phi$ solution of
(\ref{poldef}) can be expressed in
$\mathcal{H}^{(N_a, N_b)}$ as
\begin{eqnarray}
\hat{E}^{13}_\Phi & = & |1; N_a, N_b -1 \rangle
\langle 3; N_a-1, N_b -1 | \nonumber \\
& - & | 3; N_a - 1, N_b -1 \rangle \langle 1; N_a, N_b -1 |
\nonumber \\
& + & |2; N_a - 1, N_b \rangle \langle 2; N_a - 1, N_b | .
\end{eqnarray}
As one would expect, it acts as a ladder-like operator
\begin{eqnarray}
& \hat{E}^{13}_\Phi |3; N_a - 1, N_b -1 \rangle =
|1; N_a, N_b -1 \rangle , &  \nonumber \\
& & \\
& \hat{E}^{13}_\Phi |2; N_a-1, N_b \rangle =
|2; N_a -1, N_b \rangle , &
\nonumber
\end {eqnarray}
and thus has eigenvectors
\begin{eqnarray}
\label{eig13}
| \Phi^{13}_0 \rangle & = &
| 2; N_a-1, N_b \rangle ,
\nonumber \\
& & \\
| \Phi^{13}_\pm \rangle & = &
\frac{1}{\sqrt{2}}
( | 3,  N_a-1, N_b -1 \rangle
\pm i  | 1;  N_a, N_b -1 \rangle ) ,
\nonumber
\end{eqnarray}
while the eigenvalues of $\hat{\Phi}^{13}$ are
$0$ and $\pm \pi/2$, respectively.

Obviously, a similar reasoning for the transition
$2 \leftrightarrow 3$ gives the corresponding
operator $\hat{E}^{23}_\Phi$ as
\begin{eqnarray}
\hat{E}^{23}_\Phi & = & |2; N_a -1 , N_b  \rangle
\langle 3; N_a-1, N_b -1 | \nonumber \\
& - & | 3; N_a - 1, N_b -1 \rangle \langle 2; N_a-1 , N_b |
\nonumber \\
& + & |1; N_a, N_b - 1 \rangle \langle 1; N_a, N_b-1 | ,
\end{eqnarray}
with eigenvectors
\begin{eqnarray}
\label{eig23}
| \Phi^{23}_0 \rangle & = & | 1;  N_a, N_b -1 \rangle ,
\nonumber \\
& & \\
| \Phi^{23}_\pm \rangle & = &
\frac{1}{\sqrt{2}}
( | 3,  N_a-1, N_b -1 \rangle \pm i  | 2;  N_a-1, N_b \rangle ) ,
\nonumber
\end{eqnarray}
and the same eigenvalues as before. The states
(\ref{eig13}) and (\ref{eig23}) are the basis for
our subsequent analysis of the dynamics of the
relative phase.

\subsection{Relative-phase distribution function}

For any state, the information one can reap using
a measurement of some observable is given by the
statistical distribution of the measurement outcomes.
Once again, let us first focus on the relative phase
between the field mode $a$ and the dipole transition
$1 \leftrightarrow 3$. According to Eq.~(\ref{Ptheta})
it seems natural to define the probability
distribution function of a state described by
the density matrix $\hat{\varrho}(t)$ as
\begin{equation}
P(N_a, N_b, \Phi_r^{13}, t) =
\mathrm{Tr} [\hat{\varrho}(t) \
 | \Phi_r^{13} \rangle \langle \Phi_r^{13}| ]  ,
\end{equation}
where the vectors $| \Phi_r^{13} \rangle$ are given
in Eq.~(\ref{eig13}) and the subscript $r$ runs
the three possible eigenvalues 0, and $\pm \pi/2$.
This expression can be interpreted as a joint
probability distribution for the relative phase
and the excitation operators $\hat{N}_a$ and
$\hat{N}_b$. From this function, we can derive
the distribution for the relative phase as the
marginal distribution
\begin{equation}
P(\Phi_r^{13}, t) =
\sum_{N_a, N_b = 0}^\infty P(N_a, N_b, \Phi_r^{13}, t) .
\end{equation}
For a general state as in Eq.~(\ref{Psit}), one has
\begin{equation}
P(N_a, N_b, \Phi_r^{13}, t) =
| \langle \Phi_r^{13}| \Psi(t) \rangle |^2 ,
\end{equation}
which, through direct calculation, gives
\begin{eqnarray}
\label{Pf13}
P(\Phi_0^{13}, t) & = &
\sum_{N_a, N_b = 0}^\infty
\left |
\sum_{j=1}^3 Q_{N_a - \mu_j} Q_{N_b - \nu_j} \
c_j \ {\mathcal{U}}_{2j}^{(N_a, N_b)} (t) \right |^2 ,
\nonumber \\
& & \\
P(\Phi_\pm^{13}, t) & = &
\sum_{N_a, N_b = 0}^\infty
\left |
\sum_{j=1}^3 Q_{N_a - \mu_j} Q_{N_b - \nu_j} \
c_j \ [ {\mathcal{U}}_{3j}^{(N_a, N_b)} (t)
\pm i {\mathcal{U}}_{1j}^{(N_a, N_b)} (t) ]
\right |^2 .
\nonumber
\end{eqnarray}
Much in the same way one also gets analogous
results for the transition $2 \leftrightarrow 3$:
\begin{eqnarray}
\label{P23}
P(\Phi_0^{23}, t) & = &
\sum_{N_a, N_b = 0}^\infty
\left |
\sum_{j=1}^3 Q_{N_a - \mu_j} Q_{N_b - \nu_j} \
c_j \ {\mathcal{U}}_{1j}^{(N_a, N_b)} (t) \right |^2 ,
\nonumber \\
& & \\
P(\Phi_\pm^{23}, t) & = &
\sum_{N_a, N_b = 0}^\infty
\left |
\sum_{j=1}^3 Q_{N_a - \mu_j} Q_{N_b - \nu_j} \
c_j  \ [ {\mathcal{U}}_{3j}^{(N_a, N_b)} (t)
\pm i {\mathcal{U}}_{2j}^{(N_a, N_b)} (t) ]
\right |^2 .  \nonumber
\end{eqnarray}
This is our basic and compact result to analyze the
evolution of the relative phase.

We have numerically evaluated this distribution
for the three allowed values of the relative phase
for the case when the atom is initially in the
ground state $| 1 \rangle$ and modes $a$ and
$b$ are in a coherent state with a mean number of
photons $\bar{n}_a$ and $\bar{n}_b$, respectively.
For computational simplicity  we have used the
rescaled time
\begin{equation}
\tau = \frac{g_a t}{2 \pi \sqrt{\bar{n}_a}} ,
\end{equation}
in all the plots and have assumed that $g_a = g_b$,
which is not a serious restriction.

In Fig.~2 we have plotted a typical situation of a
weak field, in which the number of excitations in
the system is small, say $\bar{n}_a \sim \bar{n}_b
\sim 1$. The pattern shows an almost oscillatory
behavior, which can be easily understood if we retain
only the two first terms in the sums over $N_a$
and $N_b$ and use the explicit form of the
evolution operator $\mathcal{U}$.  A relevant and
general feature that is apparent from this figure is that the
probabilities associated with $\Phi^{ij}_+$ and
$\Phi^{ij}_-$ always oscillate out of phase, a
point previously demonstrated for the case of
the Jaynes-Cummings model~\cite{Luis97}.

Perhaps more interesting is the case of strong-field
dynamics,  when the number of excitations in the
system is large and so $\bar{n}_a$ or $\bar{n}_b$,
or both, are large. In Fig.~3 we have plotted the
relative-phase probabilities for  $\bar{n}_a= 50$
and  (a) $\bar{n}_b = 0.5$, and (b) $\bar{n}_b = 50$
photons, with the atom  initially in the ground
state $| 1 \rangle$. When $\bar{n}_b=0.5$, the
distribution $P(\Phi_0^{13}, t)$ (which is the
probability of finding the atom in the
level $| 2 \rangle$) is almost negligible, while
$P(\Phi_\pm^{13}, t)$ show collapses and revivals.
One may interpret this physically as follows:
the transition $1 \leftrightarrow 3$ is so intense
due to stimulated processes in mode $a$ that there is no
population transfer to level $| 2 \rangle$, which
originates a regular oscillation of the dipole
$1 \leftrightarrow 3$ with the corresponding
collapses and revivals in the relative phase.
The well-known (nearly) time-independent behavior
in the time windows between collapse and revival
is also clear.  The probability of finding the
atom in the level $| 1 \rangle$, $P(\Phi_0^{23}, t)$,
tends to be 1/2 (except at the revivals), which
confirms that the transition $1 \leftrightarrow 3$ is
almost saturated.

When $\bar{n}_b$ grows, the position of the
collapses and revivals changes, according to
standard estimates~\cite{Eberly85}. When
$\bar{n}_a = \bar{n}_b = 50$, $P(\Phi_0^{13}, t)$
is centered at 1/4, while $P(\Phi_0^{23}, t)$
is centered at 1/2, showing that the populations
tend to be equidistributed because now the
transition $2 \leftrightarrow 3$ is almost
saturated too.

Very interesting physical phenomena arise when
one considers coherent superpositions of atomic
states, because it is then possible to cancel
absorption or emission under certain conditions,
i.e., the atom is effectively transparent to the
incident field even in the presence of resonant
transitions. A semiclassical analysis~\cite{Scully99},
in which the fields are treated as $c$-numbers
and described by the complex Rabi frequencies
$\Omega_a e^{- i \theta_a}$ and
$\Omega_b e^{- i \theta_b}$ (note that $\theta_a$
and $\theta_b$ are the `phases' of the respective
fields), easily shows that when the initial atomic
state is a superposition of the two lower levels of
the form
\begin{equation}
\label{trap}
| \Psi (0) \rangle_{\mathrm{A}} =
\frac{1}{\sqrt{2}} ( | 1 \rangle + e^{i \varphi} \ | 2 \rangle ) ,
\end{equation}
coherent trapping occurs whenever
\begin{equation}
\Omega_a = \Omega_b ,
\qquad
\theta_a - \theta_b - \varphi = \pm \pi .
\end{equation}
In other words, when these conditions are
fulfilled the population is trapped in the
lower states and there is no absorption.

To corroborate this behavior valid in the
strong-field limit, let us note that, although the
transition $1 \leftrightarrow 2$ is dynamically
forbidden, one can still define phase eigenvectors
for it:
\begin{eqnarray}
\label{eig12}
| \Phi^{12}_0 \rangle & = & | 3, N_a-1, N_b-1 \rangle ,
\nonumber \\
& & \\
| \Phi^{12}_\pm \rangle & = &
\frac{1}{\sqrt{2}}
( | 2;  N_a-1, N_b  \rangle \pm i  | 1;  N_a, N_b -1 \rangle ) .
\nonumber
\end{eqnarray}
In Fig.~4 we have plotted the probabilities $P(\Phi_r^{12}, t)$
when the atom is initially in a trapped state like (\ref{trap})
with $\bar{n}_a = \bar{n}_b = 50$. We see that
$P(\Phi_0^{12}, t)$,  which is the probability of
finding the atom in the upper state, shows the remarkable
behavior of trapping, except for the presence of very small
superimposed oscillations. Note also that $P(\Phi_\pm^{12}, t)$
also shows the same kind of behavior.

\section{Conclusions}

In this paper we have investigated an appropriate
operator for the quantum description of the relative
phase in the interaction of a three-level atom with
quantum fields. To this end, we have resorted to a
proper polar decomposition of the corresponding
amplitudes, which has been justified on physical
grounds as well as using the theory of polynomial
deformations of su(3).

From the phase states obtained in this procedure
we have defined a probability distribution function
for the relative phase and studied its time evolution,
showing how the formalism could be applied to
understanding more involved phenomena.

\newpage

\begin{figure}
\caption{The energy scheme of a three-level
$\Lambda$-type atom interacting with two
single-mode quantum fields, coupling the two
ground states to a common excited atomic state.}
\end{figure}

\begin{figure}
\caption{The probability distribution function
for the six allowed values of the relative phase
as a function of the rescaled time $\tau$ in
the case of a weak field with $\bar{n}_a =
\bar{n}_b = 1$.}
\end{figure}

\begin{figure}
\caption{The probability distribution function
for the six allowed values of the relative phase
as a function of the rescaled time $\tau$ in
the strong-field limit with: a) $\bar{n}_a = 50$,
$\bar{n}_b = 0.5$ and b) $\bar{n}_a = 50$,
$\bar{n}_b = 50$.}
\end{figure}

\begin{figure}
\caption{The probability distribution function
for the allowed values of the relative phase
$\hat{\Phi}^{12}$ as a function of the rescaled
time $\tau$ for a trapped state with
$\bar{n}_a = \bar{n}_b = 50$.}
\end{figure}

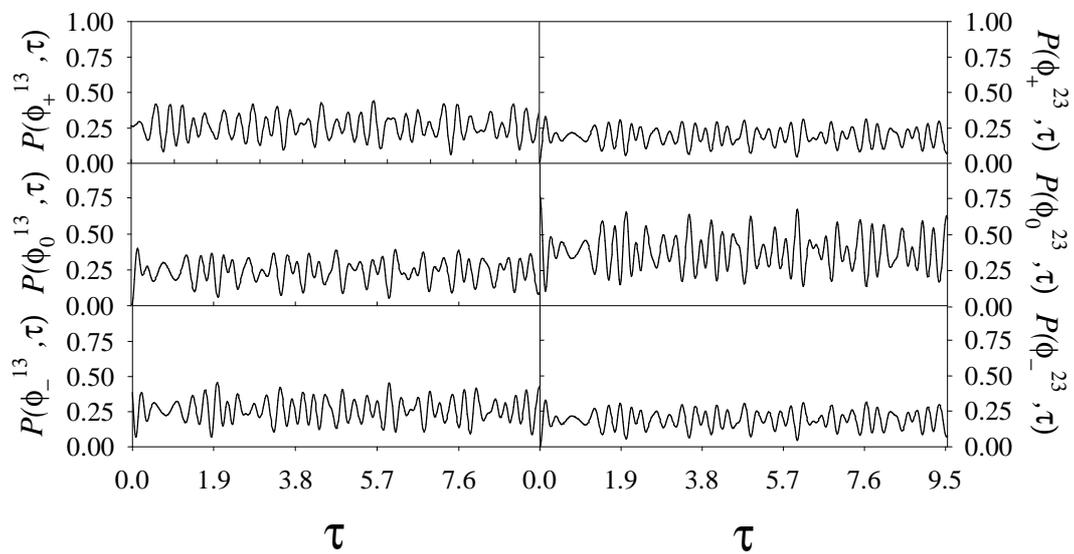
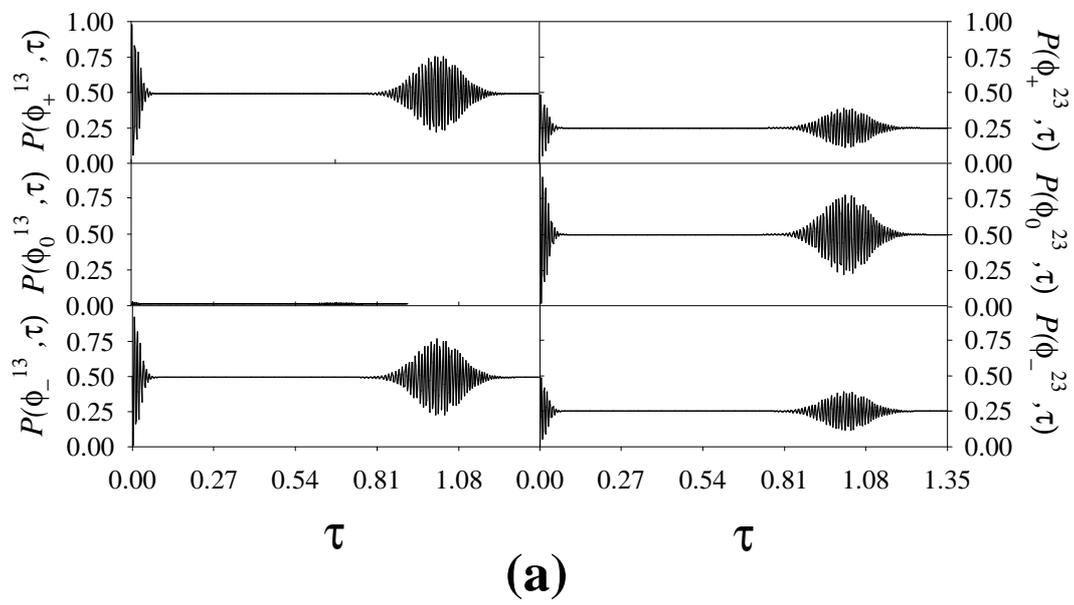
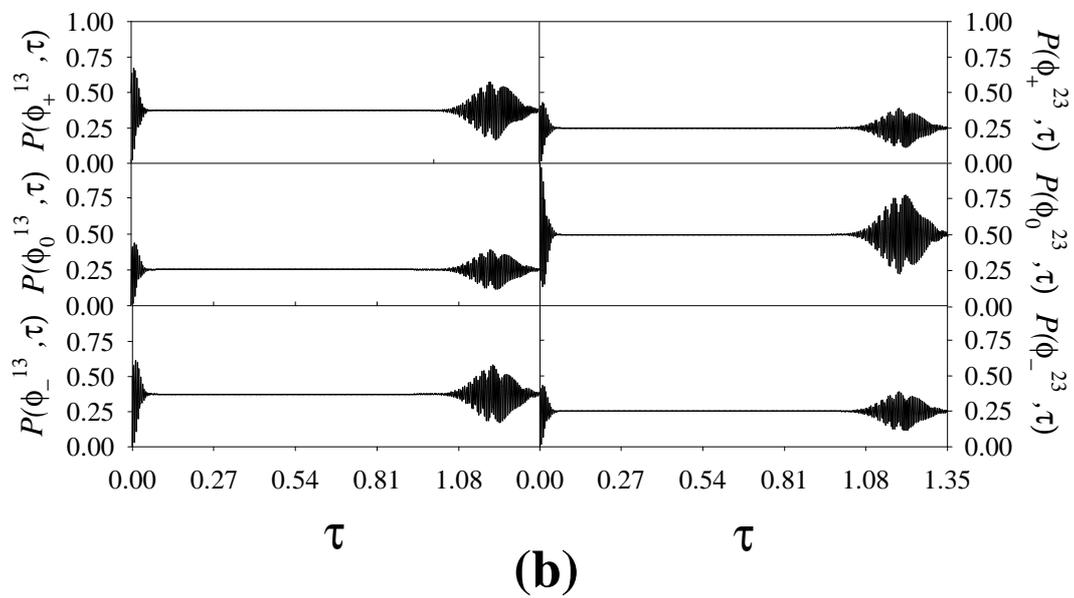
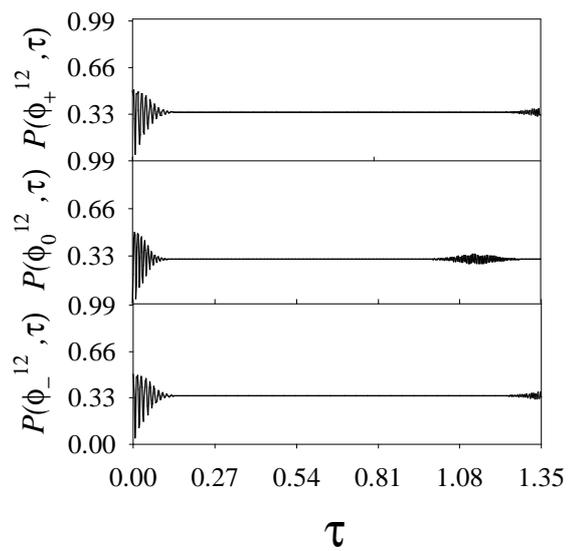
\end{document}